\documentclass[12pt]{article}
\date{August 16, 2017}
\usepackage[usenames,dvipsnames]{color}
\usepackage{amsmath,amsthm,amsfonts,amssymb,multicol,amscd,amsbsy}
\usepackage{graphicx}
\usepackage{booktabs}
\usepackage{array}
\usepackage{subfigure}
\usepackage{enumerate}
\usepackage{url}
\usepackage[nodayofweek]{datetime}
\usepackage[english]{babel}
\usepackage{mathtools}
\usepackage[toc,page]{appendix}
\usepackage{verbatim} 
\usepackage{amsthm} 
\usepackage{enumitem}
\usepackage{enumerate}
\usepackage{bbm} 
\usepackage[normalem]{ulem} 
\usepackage{bm}
\usepackage{hyperref}

\DeclarePairedDelimiter\ceil{\lceil}{\rceil}

\usepackage{fullpage}
\usepackage{authblk}

\newcommand{\be}{\begin}
\newcommand{\e}{\end}
\newcommand{\beq}{\begin{equation}}
\newcommand{\eeq}{\end{equation}}
\newcommand{\beqs}{\begin{equation*}}
\newcommand{\eeqs}{\end{equation*}}
\newcommand{\bal}{\begin{align}}
\newcommand{\eal}{\end{align}}
\newcommand{\bals}{\begin{align*}}
\newcommand{\eals}{\end{align*}}

\newcommand{\ol}{\overline}

\renewcommand{\l}{\left}
\renewcommand{\r}{\right}

\renewcommand{\d}{\mathrm{d}} 

\newcommand{\set}[1]{\mathbb{#1}}

\newcommand{\curly}[1]{\mathcal{#1}}

\newcommand{\R}{\set{R}}
\newcommand{\C}{\set{C}}

\newcommand{\om}{\omega}
\newcommand{\Om}{\Omega}

\newcommand{\lam}{\lambda}

\newcommand{\al}{\alpha}
\newcommand{\de}{\delta}

\newcommand{\scp}[2]{\langle#1,#2\rangle}

\newcommand{\ket}[1]{|#1\rangle}

\newcommand{\ketbra}[2]{ |#1 \rangle \langle #2|}

\renewcommand{\it}{\infty}

\newcommand{\Tr}{\mathrm{Tr}}	

\theoremstyle{definition}

\numberwithin{equation}{section}

\theoremstyle{remark}

\def\dotuline{\bgroup
  \ifdim\ULdepth=\maxdimen  
   \settodepth\ULdepth{(j}\advance\ULdepth.4pt\fi
  \markoverwith{\begingroup
  \advance\ULdepth0.08ex
  \lower\ULdepth\hbox{\kern.15em .\kern.1em}%
  \endgroup}\ULon}

\def\dashuline{\bgroup
  \ifdim\ULdepth=\maxdimen  
   \settodepth\ULdepth{(j}\advance\ULdepth.4pt\fi
  \markoverwith{\kern.15em
  \vtop{\kern\ULdepth \hrule width .3em}%
  \kern.15em}\ULon}
\allowdisplaybreaks

\begin{document}
\title{On multivariate trace inequalities of Sutter, Berta and Tomamichel}

\author{Marius Lemm\thanks{mlemm@caltech.edu} }
\affil{\small{Department of Mathematics, Caltech\\ Department of Mathematics, Harvard University}}

\maketitle

\abstract{
We consider a family of multivariate trace inequalities recently derived by Sutter, Berta and Tomamichel. These inequalities generalize the Golden-Thompson inequality and Lieb's three-matrix inequality to an arbitrary number of matrices in a way that features complex matrix powers. We show that their inequalities can be rewritten as an $n$-matrix generalization of Lieb's original three-matrix inequality. The complex matrix powers are replaced by resolvents and appropriate maximally entangled states. We expect that the technically advantageous properties of resolvents, in particular for perturbation theory, can be of use in applications of the $n$-matrix inequalities, e.g., for analyzing the rotated Petz recovery map in quantum information theory.
}

\section{Introduction and main result}

Arguably the most famous trace inequality is the Golden-Thompson inequality \cite{Golden,Thompson}, which asserts that for all positive definite matrices $A_1$ and $A_2$,
\beq\label{eq:GT}
\Tr[\exp(\log A_1+\log A_2)]\leq \Tr[A_1 A_2].
\eeq
In 1973, Lieb proved a generalization of \eqref{eq:GT} to three positive definite matrices $A_1$, $A_2$ and $A_3$, namely,
\beq\label{eq:lieb3mi}
\Tr[\exp(\log A_1 + \log A_2 + \log A_3)]\leq \Tr[A_3 T_{A_2^{-1}}(A_1)].
\eeq
The right-hand side features $T_X(Y)$, the Fr{\'e}chet derivative of the matrix logarithm
\beq\label{eq:Tdefn}
T_X(Y):=\frac{\d}{\d r}\Big\vert_{r=0}\log(X+r Y)=\int_0^\it \frac{1}{X+\tau} Y \frac{1}{X+\tau}\d\tau,
\eeq
where $X$ and $Y$ are positive definite matrices. Here and in the following, we write $\tau$ for $\tau I$, where $I$ is an identity matrix of the appropriate dimension. The expression $T_X(Y)$ can be understood as a non-commutative analog of $X^{-1} Y$. Lieb's three-matrix inequality \eqref{eq:lieb3mi} has numerous applications, in particular to entropy inequalities in quantum information theory. 

In 2016, Sutter, Berta and Tomamichel \cite{SBT} generalized Lieb's three-matrix inequality to the following \emph{multivariate trace inequality}:
\beq\label{eq:SBTkmi}
\Tr\l[\exp\l(\sum_{k=1}^n \log A_k\r)\r]\leq \int_\R \Tr[A_n A_{n-1}^{\frac{1+it}{2}} \ldots A_2^{\frac{1+it}{2}} A_1 A_2^{\frac{1-it}{2}} \ldots A_{n-1}^{\frac{1-it}{2}}] \beta(t)\d t.
\eeq
Here $A_1,\ldots,A_n$ are positive definite matrices and 
\beq\label{eq:betadefn}
\beta(t):=\frac{\pi}{2}(1+\cosh(\pi t))^{-1}
\eeq
 is an explicit probability density function. The proof of \eqref{eq:SBTkmi} is based on powerful complex interpolation techniques. (In fact, these techniques yield a stronger inequality, namely an $n$-matrix version of the Araki-Lieb-Thirring inequality for Schatten norms \cite{SBT}; see also \cite{HST}. The derivation of the $n$-matrix inequality \eqref{eq:SBTkmi} is discussed following Corollary 3.3 in \cite{SBT}.)

 The $n=4$ version of \eqref{eq:SBTkmi} has applications to recoverability questions in quantum information theory, which we will discuss later on. We mention that there exist other multivariate trace inequalities generalizing the Golden-Thompson inequality, in particular one proved by Hansen \cite{Hansen}, and we refer to \cite{SBT} for a summary of the pertinent literature.\\

Given these results, it is natural to relate the (a priori rather different looking) inequalities \eqref{eq:lieb3mi} and \eqref{eq:SBTkmi} for $n=3$. In this vein, \cite{SBT} observed that 
\beq\label{eq:SBTlemma}
\int_\R  A_2^{\frac{1+it}{2}} A_1 A_2^{\frac{1-it}{2}}  \beta(t)\d t = T_{A_2^{-1}}(A_1).
\eeq
By diagonalizing $A_2$, this is seen to be equivalent to the equality
$\sqrt{xy} \int (y/x)^{it/2}\beta(t)\d t=\int_0^\it (x+\tau)^{-1}(y+\tau)^{-1}\d\tau$, which holds for all real numbers $x,y>0$. The identity \eqref{eq:SBTlemma} implies that when $n=3$, inequality \eqref{eq:SBTkmi} is the same as Lieb's three-matrix inequality \eqref{eq:lieb3mi} since
\beq\label{eq:3misame}
\int_\R  \Tr[A_3 A_2^{\frac{1+it}{2}} A_1 A_2^{\frac{1-it}{2}}]  \beta(t)\d t = \Tr[A_3 T_{A_2^{-1}}(A_1)].
\eeq

\subsection{The main result}
The main result, Theorem \ref{thm:main1} below, shows how to generalize the form of Lieb's three-matrix inequality \eqref{eq:lieb3mi} to $n$ matrices. 

The generalization is derived from the $n$-matrix inequality \eqref{eq:SBTkmi}, by appropriately rewriting its right-hand side. Specifically, we observe that the identity \eqref{eq:3misame} can be generalized to $n$ matrices by introducing a suitable, non-trivial tensor product structure (Lemma \ref{lm:key}). 

\noindent
In preparation for the main result, Theorem \ref{thm:main1}, we introduce some notation:

\be{itemize}
\item Given $x\in\R$, we write $\ceil{x}$ for the smallest \emph{positive} integer greater or equal to $x$. Note that this means $\ceil{0}=1$. 

\item We write $\{\al_k\}_{k\geq 2}$ for the \emph{Thue-Morse sequence}. The Thue-Morse sequence is a $\{0,1\}$-valued sequence which can be defined as the substitution sequence associated to the substitution rules $0\rightarrow 01$ and $1\rightarrow 10$, with initial value $0$.

We use the following equivalent definition: We set $\al_{j+2}:=\beta_{j}$ ($j\geq 0$), where the sequence $\{\beta_j\}_{j\geq 0}$ is inductively defined by the relations
\beq\label{eq:TMdefn}
\beta_0=0,\qquad \beta_{2j}=\beta_j,\qquad \beta_{2j+1}=1-\beta_j.
\eeq
E.g., the first four elements of the Thue-Morse sequence are $(\al_2,\al_3,\al_4,\al_5)=(0,1,1,0)$. 

\item Let $\curly{H}$ be a $d$-dimensional Hilbert space, with a fixed orthonormal basis $\{\ket{l}\}_{l=1}^{d}$. Given $m\geq 1$, we write $\ket{\Om_m}$ for the maximally entangled state on $\curly{H}^{\otimes m}\otimes \curly{H}^{\otimes m}$ associated to this basis, i.e.,
$$
\ket{\Om_m}:=\sum_{l_1,\ldots,l_m=1}^{d}\ket{l_1\otimes \ldots\otimes l_m}\otimes \ket{l_1\otimes \ldots\otimes l_m}.
$$
We emphasize that $\ket{\Om_m}$ is \emph{not} normalized, i.e., $\|\Om_m\|=d^{m/2}\neq 1$. We denote the corresponding (also non-normalized) projector by 
$$
P_m=\ketbra{\Om_m}{\Om_m}.
$$
We refer e.g.\ to \cite{WildeBook,Wolff} for background on maximally entangled states. (We emphasize that we use a different normalization convention, however.)

\item We write $\ol{A_k}$ for the complex conjugate of the positive definite matrix $A_k$. By definition, the complex conjugate is taken with respect to the fixed $\{\ket{l}\}_l$ basis from above. (We mention the fact that the matrix $\ol{A_k}$ is still positive definite and that $\ol{A_k}=A_k^T$.) 

We write $C$ for the complex conjugation operator in the $\{\ket{l}\}_l$ basis. For any matrix $X$ on $\curly{H}$, it holds that
\beq\label{eq:tildeAdefn}
C^{\al_k}X C^{\al_k} :=
\be{cases}
X,\quad \textnormal{if } \al_{k}=0,\\
\ol{X},\quad \textnormal{if } \al_{k}=1.
\e{cases}
\eeq
\e{itemize}

Our main result is the following alternative, equivalent form of the multivariate trace inequality \eqref{eq:SBTkmi} from \cite{SBT}. We recall the definitions \eqref{eq:Tdefn} of $T_X(Y)$ and \eqref{eq:betadefn} of $\beta(t)$.

\be{thm}\label{thm:main1}
Let $n\geq3$ and let $A_1,\ldots,A_n$ be positive definite matrices on a finite-dimensional Hilbert space $\curly{H}$. Set $n':=\ceil{\log_2 (n-2)}$ and $\rho=2^{\ceil{\log_2(n-2)}}-(n-2)$. We have the inequality
\beq\label{eq:main1}
\begin{aligned}
&\Tr_{\curly{H}}\l[\exp\l(\sum_{k=1}^n \log A_k\r)\r]\\
\leq
&\Tr_{ \curly{H}^{\otimes 2^{n'}}}\l[P_{2^{n'-1}}T_{\bigotimes_{k=2}^{n-1} C^{\al_k}A_{k}^{-1} C^{\al_k}\otimes I_{\curly{H}}^{\otimes \rho}}\l(A_1\otimes \ol{A_n} \otimes \bigotimes_{j=0}^{n'-2} P_{2^j}\r)\r].
\end{aligned}
\eeq
\e{thm}

We may write out the matrix $T_X(Y)$ appearing in \eqref{eq:main1} as an integral by using \eqref{eq:Tdefn}. When we do so, \eqref{eq:main1} reads
$$
\begin{aligned}
&\Tr_{\curly{H}}\l[\exp\l(\sum_{k=1}^n \log A_k\r)\r]\\
\leq&\int_0^\it 
\Tr_{ \curly{H}^{\otimes 2^{n'}}}\l[P_{2^{n'-1}}\l(\frac{1}{\bigotimes_{k=2}^{n-1} C^{\al_k}A_{k}^{-1} C^{\al_k}+\tau}\otimes I_{\curly{H}}^{\otimes \rho}\r)\r.\\
&\qquad\qquad\qquad\l.\l(A_1\otimes \ol{A_n} \otimes \bigotimes_{j=0}^{n'-2} P_{2^j}\r) \l(\frac{1}{\bigotimes_{k=2}^{n-1} C^{\al_k}A_{k}^{-1} C^{\al_k}+\tau}\otimes I_{\curly{H}}^{\otimes \rho}\r)\r]\d\tau.
\end{aligned}
$$

\be{rmk}\label{rmk:main}
\be{enumerate}[label=(\roman*)]

\item The identity matrix $I_{\curly{H}}^{\otimes \rho}$ is present for dimensional reasons. We note that $\rho=0$ iff $n-2=2^m$ for some integer $m$.

\item We recall that the first four elements of the Thue-Morse sequence $\{\al_k\}_{k\geq 2}$ are $(0,1,1,0)$. Therefore, we have
$$
\bigotimes_{k=2}^5 C^{\al_k}X_{k}C^{\al_k} = X_2\otimes \ol{X_3}\otimes \ol{X_4} \otimes X_5,
$$
and this already suffices for many cases of interest. 

\item If $n=3$, then \eqref{eq:main1} is just Lieb's three-matrix inequality \eqref{eq:lieb3mi}. Indeed, in this case we have $n'=\ceil{0}=1$ per our convention and \eqref{eq:main1} says
$$
\Tr_{\curly{H}}[\log( A_1+\log A_2+\log A_3)]\leq \Tr_{\curly{H}^{\otimes 2}}\l[P_1 T_{A_2\otimes I_{\curly{H}}}(A_1\otimes \ol{A_3})\r]
= \Tr_{\curly{H}}[ A_3 T_{A_2}(A_1)].
$$
The second equality is an elementary property of the maximally entangled state; its general form is recalled in \eqref{eq:key} below. 

\item If $n=4$, then $n'=1$ and \eqref{eq:main1} gives the new inequality
\beq\label{eq:newfour}
\Tr_{\curly{H}}[\exp(\log A_1+\log A_2+\log A_3+ \log A_4)]
\leq \Tr_{\curly{H}^{\otimes 2}}\l[P_1 T_{(A_2 \otimes \ol{A_3})^{-1}}(A_1\otimes \ol{A_4})\r],
\eeq
where $P_1=\sum_{l,m}\ketbra{l \otimes l}{m\otimes m}$. The $n=4$ case is particularly relevant as it can be applied to recoverability questions in quantum information theory \cite{SBT}. We explore the potential usefulness of the new formulation \eqref{eq:newfour} in this context at the end of the introduction. 

\item
Our proof of Theorem \ref{thm:main1} relies on the $n$-matrix inequality \eqref{eq:SBTkmi} as input. Specifically, Theorem \ref{thm:main1} is proved by showing that the right-hand side of the inequality \eqref{eq:SBTkmi} can be rewritten as the right-hand side of \eqref{eq:main1}, up to a permutation of the $A_k$ (Lemma \ref{lm:key}).

The maximally entangled states arise via the following elementary property: For any two self-adjoint matrices $X$ and $Y$ on $\curly{H}^{\otimes m}$, we have
\beq\label{eq:key}
\Tr_{\curly{H}^{\otimes m}}[XY]=\Tr_{\curly{H}^{\otimes 2m}}[P_m(X\otimes Y^T)]=\Tr_{\curly{H}^{\otimes 2m}}[P_m(X\otimes \ol{Y})].
\eeq
The basic idea of the proof of Lemma \ref{lm:key} is to use this identity iteratively on the right-hand side of \eqref{eq:SBTkmi}, splitting in half the set of matrices $A_2,\ldots,A_{n-1}$ that come with a complex power at every iteration step. The iteration stops after $n'$ steps, when one obtains an expression that can be rewritten via \eqref{eq:SBTlemma}. At every step, the Hilbert space is tensored with itself as in \eqref{eq:key}, and this is what produces the projections $P_{2^j}$ in \eqref{eq:main1}.

\item By \eqref{eq:Tdefn}, we can rewrite the right-hand side of \eqref{eq:main1} in terms of the Fr{\'e}chet derivative of a matrix logarithm. Then, \eqref{eq:main1} reads
$$
\begin{aligned}
&\Tr_{\curly{H}}\l[\exp\l(\sum_{k=1}^n \log A_k\r)\r]\\
\leq &\frac{\d}{\d r}\Big\vert_{r=0}\Tr_{\curly{H}^{\otimes 2^{n'}}}
\l[P_{n'-1}\exp\l(-\log \curly{A}
+\log\l(\curly{A}+r A_1\otimes \ol{A_n} \otimes \bigotimes_{j=0}^{n'-2} P_{2^j}\r)\r)\r],
\end{aligned}
$$
where we abbreviated $\curly{A}:=\bigotimes_{k=2}^{n-1} C^{\al_k} A^{-1}_{k}C^{\al_k}\otimes I_{\curly{H}}^{\otimes \rho}$. By using homogeneity as in Lemma 5 of \cite{LiebWYD}, we see that this inequality is related to the concavity of the homogeneous trace function
$$
K\mapsto \Tr_{\curly{H}^{\otimes 2^{n'}}}[P_{n'-1}\exp(-\log \curly{A}+\log K)]
$$
near the matrix $K=\curly{A}$. In particular, a direct proof of the concavity of this trace function would yield a new, direct proof of Theorem \ref{thm:main1} that does not rely on \eqref{eq:SBTkmi}. 
\e{enumerate}
\e{rmk}

\subsection{A related inequality}

Notice that the right-hand side in \eqref{eq:main1} is of the general form $\Tr[P T_{X}(Y)]$ for any $n$, where $P$ is a non-negative definite matrix and $X$ and $Y$ are appropriate positive definite matrices. Therefore, for any $n$, it can be estimated from below via Lieb's original three-matrix inequality \eqref{eq:3misame}, up to an approximation argument to make $P$ \emph{strictly} positive definite. This procedure yields a different inequality than \eqref{eq:main1} which we state in Proposition \ref{prop:related} below, in the case $n=4$. (The result generalizes to higher $n$.)

The obtained inequality \eqref{eq:prop} appears to be weaker than the four-matrix inequality \eqref{eq:newfour}, but its proof only requires Lieb's three-matrix inequality \eqref{eq:lieb3mi} as input.

\be{prop}\label{prop:related}
Let $A_1,A_2,A_3$ and $A_4$ be positive definite matrices on a $d$-dimensional Hilbert space $\curly{H}$. Then we have
\beq\label{eq:prop}
\begin{aligned}
d \exp\l(\frac{1}{d}\Tr[\log A_1+\log A_2+\log A_3+\log A_4]\r)
\leq \Tr_{\curly{H}^{\otimes 2}}\l[P_1 T_{(A_2 \otimes \ol{A_3})^{-1}}(A_1\otimes \ol{A_4})\r].
\end{aligned}
\eeq
\e{prop}

The statement and proof of Proposition \ref{prop:related} generalize to other values of $n>4$.

\be{rmk}\label{rmk:prop}
\be{enumerate}[label=(\roman*)]
\item The right-hand side in \eqref{eq:prop} can alternatively be expressed in terms of complex matrix powers by using  \eqref{eq:foursame}.

\item While the appearance of the dimension $d$ on the left-hand side of \eqref{eq:prop} may appear unpleasant at first sight, it actually makes the expression \eqref{eq:prop} grow only linearly in $d$ when evaluated on $A_k=I_{\curly{H}}$, as opposed to exponentially. The expression  $\Tr[\exp(\sum_k \log A_k)]$ shows the same linear growth in $d$ in that case. Motivated by partition functions in statistical mechanics, we say that the left-hand side in \eqref{eq:prop} has the ``thermodynamically correct scaling''. (This  property is lost for other inequalities which exchange trace and exponential, as is exemplified by the Peierls-Bogolubov inequality.)
\e{enumerate}
\e{rmk}

\subsection{Discussion}
The main purpose of Theorem \ref{thm:main1} is to provide an $n$-matrix generalization of Lieb's original three-matrix inequality \eqref{eq:lieb3mi}.\\ 

Given that this generalization is equivalent to (and derived from) the $n$-matrix inequality of Sutter, Berta and Tomamichel \eqref{eq:SBTkmi}, one may ask if the alternative formulation \eqref{eq:main1} affords any advantages as compared to the original \eqref{eq:SBTkmi}. 

 The clear advantage of \eqref{eq:main1} is the absence of complex matrix powers. We find this aesthetically pleasing, since one is bounding a manifestly real-valued quantity $\Tr[\exp(\sum_k \log A_k)]$. More importantly, the complex matrix powers are replaced by \emph{resolvents} (e.g.\ $A_2$ appears as $(A_2^{-1}\otimes B+\tau)^{-1}$, where $\tau>0$ and $B$ is independent of $A_2$), and these have several \emph{technical advantages} for analysis, as we recall below.

A disadvantage of \eqref{eq:main1} is the tensor product structure, which increases in complexity with $n$. However, for small $n$,  the tensor product structure is still quite manageable; see e.g.\ the case $n=4$ in \eqref{eq:newfour}. 

We recall two of the technical advantages of resolvents: 
\be{enumerate}
\item Resolvents are well-suited for performing \emph{perturbation theory}, since the perturbation series has an explicit and simple form at every finite order. In particular, the Fr{\'e}chet derivative of a resolvent can be computed explicitly.
\item The \emph{commutator} with a resolvent can be written in a simple form. 
\e{enumerate}

Let us explain the potential usefulness of the latter fact further. In short, it allows to compare the $n$-matrix inequality \eqref{eq:SBTkmi} to variants in which the matrices $A_k$ have been reordered, e.g.\ to remove the complex powers. More precisely, one can bound the difference between the right-hand side in \eqref{eq:SBTkmi} and related expressions without complex powers \emph{explicitly in terms of the commutators between the different $A_k$}. As an example for this, we take $n=3$ for simplicity and rewrite the difference
$$
\begin{aligned}
& A_1 A_2 - \int_\R  A_2^{\frac{1+it}{2}} A_1 A_2^{\frac{1-it}{2}}  \beta(t)\d t\\
=& \int_0^\it \l(A_1 \l(\frac{1}{A_2^{-1}+\tau}\r)^2-\frac{1}{A_2^{-1}+\tau} A_1 \frac{1}{A_2^{-1}+\tau}\r)\d\tau\\
=& \int_0^\it \l[A_1,\frac{1}{A_2^{-1}+\tau}\r]\frac{1}{A_2^{-1}+\tau}\d\tau\\
=& \int_0^\it \frac{1}{A_2^{-1}+\tau}A_2^{-1}\l[A_1,A_2\r]A_2^{-1}\l(\frac{1}{A_2^{-1}+\tau}\r)^2\d\tau.
\end{aligned}
$$
Note that the last expression is linear in the commutator $[A_1,A_2]$.\\

The above-mentioned technical advantages of \eqref{eq:main1} are only meaningful in the context of \emph{applications} of the $n$-matrix inequalities \eqref{eq:SBTkmi} and \eqref{eq:main1}. Here we discuss as an example the application of the four-matrix inequality to \emph{quantum information theory}.\\

Motivated by previous results in this direction, \cite{BLW,FR,SFR,STH,Wilde} and especially \cite{Jungeetal}, \cite{SBT} used their new four-matrix inequality to derive a strengthened data processing inequality, with a remainder term that involves the measured relative entropy of an appropriately \emph{``rotated'' Petz recovery channel}, called $\curly{R}$ below.

From these papers, the rotated Petz recovery channel has emerged as a natural variant of the original Petz recovery channel \cite{Petz}.

To see how Theorem \ref{thm:main1} can be useful in this context, we simply replace the application of the four-matrix inequality \eqref{eq:SBTkmi} in eq.\ (54) of \cite{SBT} with the new formulation \eqref{eq:newfour}. (Equivalently, we can apply \eqref{eq:foursame} to the result of Theorem 4.1 in \cite{SBT}.) In this way, we find the following alternative to the strengthened data processing inequality from \cite{SBT}. Using the notation from \cite{SBT}, we have
\beq\label{eq:ourDPI}
\begin{aligned}
&D(\rho_{AB}\|\sigma_{AB})-D(\rho_A\|\sigma_A)\\
\geq& D_{\mathbb M}(\rho_{AB}\|\curly{R}_{\sigma_{AB},\Tr_B}(\rho_A))\\
=&\sup_{\om_{AB}>0}\l\{\Tr_{\curly{H}_{AB}}[\rho_{AB}\log \om_{AB}]+1
-\Tr_{\curly{H}_{AB}^{\otimes 2}}[P_1 T_{\sigma_{A}\otimes I_B\otimes \ol{\sigma_{AB}}^{-1}}(\rho_A\otimes I_B\otimes \ol{\om_{AB}}))]
\r\}
\end{aligned}
\eeq
The equality provides a tool to perform analysis on the rotated Petz channel $\curly{R}_{\sigma_{AB},\Tr_B}(\rho_A)$ by using \eqref{eq:Tdefn} and the properties of resolvents mentioned above. For example, one can investigate
\be{enumerate}
\item the change of the performance of the rotated Petz recovery channel under a perturbation of the involved states, say $\sigma_{AB}\to \tilde \sigma_{AB}$, and
\item the difference between the rotated Petz recovery channel and the original Petz recovery channel, i.e., the effect of removing the complex powers.
\e{enumerate}
Note that after performing analysis with resolvents, one can return to the formulation used in \cite{SBT} via \eqref{eq:foursame} and more generally Lemma \ref{lm:key}.

To summarize this example, we hope that the alternative way to write the strengthened data processing inequality \eqref{eq:ourDPI} can be of value in analyzing the rotated Petz recovery channel, which has recently emerged as an important object in quantum information theory.

\section{Proofs}

\subsection{Proof of the main result, Theorem \ref{thm:main1}}
The proof of Theorem \ref{thm:main1} is based on the following key lemma. For its statement, we recall the notation introduced before and within Theorem \ref{thm:main1}.

\be{lm}\label{lm:key}
Let $n\geq 3$. There exists a permutation $\pi$ of $\{2,\ldots,n-1\}$ such that the following holds. Let $X_1,\ldots,X_n$ be non-negative definite matrices on a finite-dimensional Hilbert space $\curly{H}$. Then we have
\beq\label{eq:keylemma}
\begin{aligned}
&\Tr_{\curly{H}}\l[X_{n} X_{n-1}^{\frac{1+it}{2}} \ldots X_2^{\frac{1+it}{2}} X_1 X_2^{\frac{1-it}{2}} \ldots X_{n-1}^{\frac{1-it}{2}}\r]\\
=&\Tr_{ \curly{H}^{\otimes 2^{n'}}}\l[P_{2^{n'-1}}
\l(\l(\bigotimes_{k=2}^{n-1} C^{\al_k}X_{\pi(k)}C^{\al_k}\r)^{\frac{1+it}{2}}\otimes I_{\curly{H}^{\otimes \rho}}\r)
\l(X_1\otimes \ol{X_n} \otimes \bigotimes_{j=0}^{n'-2} P_{2^j}\r)\r.\\
&\qquad\qquad\qquad\;
\l.\l(\l(\bigotimes_{k=2}^{n-1} C^{\al_k}X_{\pi(k)}C^{\al_k}\r)^{\frac{1-it}{2}}\otimes I_{\curly{H}^{\otimes \rho}}\r)
\r]
\end{aligned}
\eeq
for every $t\in \R$.
\e{lm}

\be{rmk}
The permutation $\pi$ is explicit, but of no further interest to us. Here we just mention that $\pi=\mathrm{id}$ when $n=4$, i.e.,
\beq\label{eq:foursame}
\begin{aligned}
\int_\R  \Tr[A_4 A_3^{\frac{1+it}{2}} A_2^{\frac{1+it}{2}} A_1 A_2^{\frac{1-it}{2}} A_3^{\frac{1-it}{2}} ]  \beta(t)\d t
=\Tr_{\curly{H}^{\otimes 2}}\l[P_1 T_{(A_2 \otimes \ol{A_3})^{-1}}(A_1\otimes \ol{A_4})\r],
\end{aligned}
\eeq
and $\pi\neq \mathrm{id}$ when $n>4$.
\e{rmk}

We will prove Lemma \ref{lm:key} below, by induction. Assuming that it holds, we can give the

\be{proof}[Proof of Theorem \ref{thm:main1}]
Fix $n\geq 3$ and let $\pi$ be the corresponding permutation from Lemma \ref{lm:key}. The idea is to apply the inequality \eqref{eq:SBTkmi} from \cite{SBT} to a reordered collection of $\{A_k\}$, namely to
$$
X_k:=
\be{cases}
A_k,\qquad\;\;\;\textnormal{if } k=1 \textnormal{ or } k=n,\\ 
A_{\pi^{-1}(k)},\quad \textnormal{if } 2\leq k\leq n-1.
\e{cases}
$$
We first apply \eqref{eq:SBTkmi} and then Lemma \ref{lm:key} to find
$$
\begin{aligned}
&\Tr\l[\exp\l(\sum_{k=1}^n \log A_k\r)\r]=\Tr\l[\exp\l(\sum_{k=1}^n \log X_k\r)\r]\\
\leq &\int_\R \Tr[X_{n} X_{n-1}^{\frac{1+it}{2}} \ldots X_2^{\frac{1+it}{2}} X_{1} X_2^{\frac{1-it}{2}} \ldots X_{n-1}^{\frac{1-it}{2}}] \beta(t)\d t\\
=&
\int_\R
\Tr_{ \curly{H}^{\otimes 2^{n'}}}\l[P_{2^{n'-1}}
\l(\l(\bigotimes_{k=2}^{n-1} C^{\al_k}A_kC^{\al_k}\r)^{\frac{1+it}{2}}\otimes I_{\curly{H}^{\otimes \rho}}\r)
\l(A_1\otimes \ol{A_n} \otimes \bigotimes_{j=0}^{n'-2} P_{2^j}\r)\r.\\
&\qquad\qquad\qquad\qquad
\l.\l(\l(\bigotimes_{k=2}^{n-1} C^{\al_k}A_kC^{\al_k}\r)^{\frac{1-it}{2}}\otimes I_{\curly{H}^{\otimes \rho}}\r)
\r] \beta(t)\d t.
\end{aligned}
$$
In the last step, we used the equality $X_{\pi(k)}=A_k$ for $2\leq k\leq n-1$. Theorem \ref{thm:main1} now follows from the general fact that $Y^{\frac{1\pm it}{2}}\otimes I=(Y\otimes I)^{\frac{1\pm it}{2}}$, where $Y$ is any positive matrix, and the equality \eqref{eq:3misame}.
\e{proof}

It remains to give the

\be{proof}[Proof of Lemma \ref{lm:key}]

\dashuline{Step 1.}
We first show that it suffices to prove the claim for all $n$ of the form $n=2^N+2$ for some $N\geq 0$ (i.e., those with $\rho=0$). 

Suppose that the claim holds for all $n$ of the form $n=2^N+2$. Consider any $n_0$ not of this form, i.e., any $n_0$ satisfying $2^{N_0}<n_0-2<2^{N_0+1}$ for some $N_0\geq0$. Since the claim holds for $n=2^{N_0+1}+2$, we get a corresponding permutation $\pi$ of $\{2,\ldots,2^{N_0+1}+1\}$. We choose matrices $\{Y_k\}_{k=1}^{2^{N_0+1}+2}$ as follows
$$
Y_k:=
\be{cases}
X_1,\quad\; &\textnormal{if } k=1,\\
X_{n_0},\quad &\textnormal{if } k=2^{N_0+1}+2,\\
X_k,\quad\; &\textnormal{if } 2\leq \pi(k)\leq n_0-1,\\
I_{\curly{H}},\quad\;\, &\textnormal{if } n_0+1\leq \pi(k)\leq 2^{N_0+1}+1.
\e{cases}
$$
By assumption, we have the analogue of \eqref{eq:keylemma} for the $\{Y_k\}$. Using $n'=N_0+1$, this means
$$
\begin{aligned}
&\Tr[Y_{2^{N_0+1}+2} Y_{2^{N_0+1}+1}^{\frac{1+it}{2}}\ldots Y_{2}^{\frac{1+it}{2}} Y_1 Y_{2}^{\frac{1-it}{2}}\ldots Y_{2^{N_0+1}+1}^{\frac{1-it}{2}}]\\
=&\Tr_{ \curly{H}^{\otimes 2^{N_0+1}}}\l[P_{2^{N_0}}
\l(\bigotimes_{k=2}^{2^{N_0+1}+1} C^{\al_k}Y_{\pi(k)}C^{\al_k}\r)^{\frac{1+it}{2}}
\l(Y_1\otimes \ol{Y_{2^{N_0}+2}} \otimes \bigotimes_{j=0}^{N_0-1} P_{2^j}\r)\r.\\
&\qquad\qquad\qquad\;
\l.\l(\bigotimes_{k=2}^{2^{N_0+1}+1} C^{\al_k}Y_{\pi(k)}C^{\al_k}\r)^{\frac{1-it}{2}}\r].
\end{aligned}
$$
Thanks to the definitions of the $\{Y_k\}$ and $\rho=2^{N_0+1}-n_0+2$, this is simply the claimed equation \eqref{eq:keylemma} for the $\{X_k\}_{k=1}^{n_0}$ and an appropriate permutation. This concludes step 1.\\

\dashuline{Step 2.} It remains to prove Lemma \ref{lm:key} for all $n$ of the form $n=2^N+2$. To this end, we perform an induction in $N$. 

The \emph{induction base} corresponds to $N=0$, or $n=3$. It follows by combining the identity \eqref{eq:SBTlemma}, which was already proved in \cite{SBT}, with the equality in Remark \ref{rmk:main} (iii).\\ 

We come to the \emph{induction step}. Fix $N\geq 0$. We assume that Lemma \ref{lm:key} holds for $n=2^N+2$, on all finite-dimensional Hilbert spaces and for all choices of orthonormal bases defining a maximally entangled state. We fix an arbitrary finite-dimensional Hilbert space $\curly{H}$ and an orthonormal basis $\{\ket{l}\}$. We want to show that Lemma \ref{lm:key} holds for $n_0=2^{N+1}+2$ on this Hilbert space. 

By cyclicity of the trace and the key property \eqref{eq:key} of maximally entangled states, we have for every $t\in \R$,
$$
\begin{aligned}
&\Tr_{\curly{H}}[X_{n_0} X_{n_0-1}^{\frac{1+it}{2}} \ldots X_2^{\frac{1+it}{2}} X_1 X_2^{\frac{1-it}{2}} \ldots X_{n_0-1}^{\frac{1-it}{2}}]\\
=&\Tr_{\curly{H}^{\otimes 2}}\l[P_1 \l(X_{\frac{n_0}{2}}^{\frac{1+it}{2}} \ldots  X_2^{\frac{1+it}{2}} X_1 X_2^{\frac{1-it}{2}} \ldots X_{\frac{n_0}{2}}^{\frac{1-it}{2}}\otimes \ol{X_{\frac{n_0}{2}+1}^{\frac{1-it}{2}} \ldots  X_{n_0-1}^{\frac{1-it}{2}} X_{n_0} X_{n_0-1}^{\frac{1+it}{2}} \ldots X_{\frac{n_0}{2}+1}^{\frac{1+it}{2}}}\r)\r].
\end{aligned} 
$$
Using the fact that $\ol{X_k^{\frac{1\pm it}{2}}}=\ol{X_k}^{\frac{1\mp it}{2}}$, as well as some general rules for tensor products of matrices, we can rewrite this expression as follows:
$$
\begin{aligned}
&\Tr_{\curly{H}^{\otimes 2}}\l[P_1 \l(X_{\frac{n_0}{2}}^{\frac{1+it}{2}} \ldots  X_2^{\frac{1+it}{2}} X_1 X_2^{\frac{1-it}{2}} \ldots X_{\frac{n_0}{2}}^{\frac{1-it}{2}}\otimes \ol{X_{\frac{n_0}{2}+1}^{\frac{1-it}{2}} \ldots  X_{n_0-1}^{\frac{1-it}{2}} X_{n_0} X_{n_0-1}^{\frac{1+it}{2}} \ldots X_{\frac{n_0}{2}+1}^{\frac{1+it}{2}}}\r)\r]\\
=&\Tr_{\curly{H}^{\otimes 2}}\l[P_1
\l(X_{\frac{n_0}{2}}^{\frac{1+it}{2}}\otimes \ol{X_{\frac{n_0}{2}+1}}^{\frac{1+it}{2}}\r)\ldots
\l(X_2^{\frac{1+it}{2}}\otimes \ol{X_{n_0-1}}^{\frac{1+it}{2}}\r)\l(X_1\otimes \ol{X_{n_0}}\r) \r.\\
&\qquad\qquad\,\l.
\l(X_2^{\frac{1-it}{2}} \otimes \ol{X_{n_0-1}}^{\frac{1-it}{2}}\r)
\ldots\l(X_{\frac{n_0}{2}}^{\frac{1-it}{2}}\otimes \ol{X_{\frac{n_0}{2}+1}}^{\frac{1-it}{2}}\r)
\r].\\
=&\Tr_{\curly{H}^{\otimes 2}}\l[P_1
\l(X_{\frac{n_0}{2}}\otimes \ol{X_{\frac{n_0}{2}+1}}\r)^{\frac{1+it}{2}}\ldots
\l(X_2\otimes \ol{X_{n_0-1}}\r)^{\frac{1+it}{2}}\l(X_1\otimes \ol{X_{n_0}}\r) \r.\\
&\qquad\qquad\,\,\l.
\Big(X_2 \otimes \ol{X_{n_0-1}}\Big)^{\frac{1-it}{2}}
\ldots\l(X_{\frac{n_0}{2}}\otimes \ol{X_{\frac{n_0}{2}+1}}\r)^{\frac{1-it}{2}}
\r].
\end{aligned} 
 $$
Now we are in a position to apply the \emph{induction hypothesis} for $n=2^N+2=n_0/2+1$ on the finite-dimensional Hilbert space $\curly{H}^{\otimes 2}$ equipped with the orthonormal basis $\{\ket{l\otimes l'}\}$. We define
\beq\label{eq:Zdefn}
 Z_k:=X_k\otimes \ol{X_{n_0+1-k}}, \qquad 1\leq k \leq n_0/2.
\eeq
We set $n_0':=\ceil{\log_2(n_0-2)}=N+1$. The induction hypothesis yields a permutation, call it $\sigma$, of $\{2,\ldots,n_0/2\}$ such that 
 \beq\label{eq:IH}
 \begin{aligned}
&\Tr_{\curly{H}}[X_{n_0} X_{n_0-1}^{\frac{1+it}{2}} \ldots X_2^{\frac{1+it}{2}} X_1 X_2^{\frac{1-it}{2}} \ldots X_{n_0-1}^{\frac{1-it}{2}}]\\
=&\Tr_{\curly{H}^{\otimes 2}}\l[P_1
Z_{\frac{n_0}{2}}^{\frac{1+it}{2}}\ldots
Z_2^{\frac{1+it}{2}}Z_1 Z_2^{\frac{1-it}{2}}
\ldots Z_{\frac{n_0}{2}}^{\frac{1-it}{2}}
\r]\\
=&
\Tr_{ \curly{H}^{\otimes 2^{n_0'}}}\l[P_{2^{n_0'-1}} \l(\bigotimes_{l=2}^{n_0/2} C^{\al_l}Z_{\sigma(l)}C^{\al_l}\r)^{\frac{1+it}{2}}\l(Z_1\otimes P_1 \otimes \bigotimes_{j=1}^{n_0'-2} P_{2^j}\r) \l(\bigotimes_{l=2}^{n_0/2} C^{\al_l}Z_{\sigma(l)} C^{\al_l}\r)^{\frac{1-it}{2}}\r].
\end{aligned} 
 \eeq
To arrive at the last line, we used that we can identify the (non-normalized) projectors $P_{2^j}(\{\ket{l\otimes l'}\})=P_{2^{j+1}}(\{\ket{l}\})$ (where we added the dependence of the projector on the underlying basis in parentheses for emphasis). This observation allows us to replace all the projectors arising via the induction hypothesis, which are associated to the basis $\{\ket{l\otimes l'}\}$, so that \eqref{eq:IH} only features projectors associated to the basis $\{\ket{l}\}$. We also used $n'=N=n_0'-1$ and the fact that $\ol{P_1}=P_1$, which holds because complex conjugation is defined in the eigenbasis of $P_1$.\\

Given \eqref{eq:IH}, the claim \eqref{eq:keylemma} will follow once we prove 

\be{lm}\label{lm:remains}
There exists a permutation $\pi$ of $\{2,\ldots, n_0-1\}$ such that
\beq\label{eq:keylmremains}
\bigotimes_{l=2}^{n_0/2} C^{\al_l}Z_{\sigma(l)}C^{\al_l}=\bigotimes_{k=2}^{n_0-1} C^{\al_k}X_{\pi(k)}C^{\al_k}.
\eeq
\e{lm}

We now prove this lemma. For convenience, we rewrite the claim \eqref{eq:keylmremains} in terms of the shifted Thue-Morse sequence $\beta_{j}:=\al_{j+2}$ ($j\geq 0$) and the shifted permutations $\tilde \sigma$ and $\tilde \pi$ defined by
$$
\tilde \sigma(j):=\sigma(j+2),\qquad \tilde\pi(j):=\pi(j+2).
$$
Using these definitions and recalling \eqref{eq:Zdefn}, we can write the claim \eqref{eq:keylmremains} as
\beq\label{eq:keylmremains'}
\bigotimes_{j=0}^{n_0/2-2} C^{\beta_j}(X_{\tilde\sigma(j)}\otimes \ol{X_{n_0+1-\tilde\sigma(j)}})C^{\beta_j}=\bigotimes_{m=0}^{n_0-3} C^{\beta_m}X_{\tilde\pi(m)}C^{\beta_m},
\eeq
for an appropriately chosen bijection $\tilde\pi:\{0,\ldots,n_0-3\}\to \{2,\ldots,n_0-1\}$.

We recall that $\tilde\sigma$ is a bijection onto $\{2,\ldots,n_0/2\}$. We see that each $X_k$ from the set $\{X_k\}_{k=2}^{n_0-1}$ appears precisely once on both sides of \eqref{eq:keylmremains'}. Hence, we can simply read off the appropriate definition of $\tilde\pi$ (and therefore of $\pi$) from \eqref{eq:keylmremains'}. That is, we take
$$
\tilde\pi(m):=
\be{cases}
\tilde\sigma\l(\frac{m}{2}\r), &\textnormal{if } m \textnormal{ is even},\\
n_0+1-\tilde\sigma\l(\frac{m-1}{2}\r),\qquad &\textnormal{if } m \textnormal{ is odd}.
\e{cases}
$$
With this definition, the claim \eqref{eq:keylmremains'} is equivalent to
$$
\bigotimes_{j=0}^{n_0/2-2} C^{\beta_j}(X_{\tilde\sigma(j)}\otimes \ol{X_{n_0+1-\tilde\sigma(j)}})C^{\beta_j}
=\bigotimes_{j=0}^{n_0/2-2} 
\l(C^{\beta_{2j}} X_{\tilde\sigma(j)} C^{\beta_{2j}} \otimes C^{\beta_{2j+1}} X_{n_0+1-\tilde\sigma(j)} C^{\beta_{2j+1}}\r)
$$
We can now conclude this equality from the relation \eqref{eq:TMdefn}, observing in particular that it yields
$$
C^{\beta_{2j+1}} X_{n_0+1-\tilde\sigma(j)} C^{\beta_{2j+1}}
=C^{1-\beta_{j}} X_{n_0+1-\tilde\sigma(j)} C^{1-\beta_{j}}
=C^{\beta_{j}} \ol{X_{n_0+1-\tilde\sigma(j)}} C^{\beta_{j}}.
$$ 
This proves Lemma \ref{lm:remains} and thus finishes step 2 of the proof of Lemma \ref{lm:key}.
\e{proof}

\subsection{Proof of Proposition \ref{prop:related}}
We begin from the right-hand side of \eqref{eq:prop}. We introduce the normalized projectors
$$
\tilde P_1:=d^{-1} P_1,\qquad \tilde P_1^\perp:=I_{\curly{H}^{\otimes 2}}-\tilde P_1,
$$
which are true projection operators in the sense that their eigenvalues are zero and one. We approximate $\tilde P_1$ by the strictly positive definite matrix
$$
\tilde P_\delta:=\tilde P_1+\delta I_{\curly{H}^{\otimes 2}},\qquad 0<\delta<1.
$$
After implementing the approximation, we can apply Lieb's three-matrix inequality \eqref{eq:lieb3mi} to get
$$
\begin{aligned}
&\Tr_{\curly{H}^{\otimes 2}}\l[P_1 T_{(A_2 \otimes \ol{A_3})^{-1}}(A_1\otimes \ol{A_4})\r]\\
=& d\, \Tr_{\curly{H}^{\otimes 2}}\l[\tilde P_1 T_{(A_2 \otimes \ol{A_3})^{-1}}(A_1\otimes \ol{A_4})\r]
= d\, \lim_{\delta\to 0} \Tr_{\curly{H}^{\otimes 2}}\l[\tilde P_\delta T_{(A_2 \otimes \ol{A_3})^{-1}}(A_1\otimes \ol{A_4})\r]\\
\geq& d\,\liminf_{\delta\to 0}\Tr_{\curly{H}^{\otimes 2}}\l[\exp\l(\log \tilde P_\de+\log(A_2 \otimes \ol{A_3})+\log(A_1\otimes \ol{A_4})\r)\r]
\end{aligned}
$$
By the spectral theorem,
$$
\log \tilde P_\delta = \log(1+\delta) \tilde P_1 +\log (\delta) \tilde P_1^\perp
\geq \log (\delta) \tilde P_1^\perp.
$$
Thus, after changing variables to $t:=-\log \delta$, we get
\beq\label{eq:beforelemma}
\begin{aligned}
&d\, \liminf_{\delta\to 0}\Tr[\exp\l(\log \tilde P_\de+\log(A_2 \otimes \ol{A_3})+\log(A_1\otimes \ol{A_4})\r)]\\
\geq& d\,\liminf_{t\to \infty}\Tr[\exp\l(-t \tilde P_1^\perp+\log(A_2 \otimes \ol{A_3})+\log(A_1\otimes \ol{A_4})\r)].
\end{aligned}
\eeq
The right-hand side can be computed explicitly via the following lemma. The lemma can be seen as an asymptotic (and therefore much simpler) version of Stahl's formula, derived in his proof of the BMV conjecture \cite{Stahl}.

\be{lm}\label{lm:stahl}
Let $A$ be a self-adjoint matrix and let $P$ be a projection so that $\ker P=\mathrm{span}\{v\}$ for a normalized vector $v$. Then
$$
\lim_{t\to\infty}\Tr[\exp(A-tP)]=\exp\scp{v}{A v}.
$$
\e{lm}

Before we prove the lemma, we observe that it yields Proposition \ref{prop:related}. Indeed, we apply the lemma to the right-hand side in \eqref{eq:beforelemma} and note that $\ker \tilde P_1^\perp=\mathrm{span}\{\ket{\Om_1}\}$. Since $\|\Om_1\|=\sqrt{d}$, we obtain
$$
\begin{aligned}
&\Tr_{\curly{H}^{\otimes 2}}\l[P_1 T_{(A_2 \otimes \ol{A_3})^{-1}}(A_1\otimes \ol{A_4})\r]\\
\geq& d\,\liminf_{t\to \infty}\Tr[\exp\l(-t \tilde P_1^\perp+\log(A_2 \otimes \ol{A_3})+\log(A_1\otimes \ol{A_4})\r)]\\
=& d\,\exp\l(\frac{1}{d}\scp{\Om_1}{\l((\log(A_2 \otimes \ol{A_3})+\log(A_1\otimes \ol{A_4})\r)\Om_1}\r)\\
=& d\,\exp\l(\frac{1}{d}\Tr[\log A_1+\log A_2+\log A_3+\log A_4]\r)
\end{aligned}
$$
In the last step, we used the elementary property \eqref{eq:key} of the maximally entangled state $\Om_1$. This proves Proposition \ref{prop:related}, assuming the validity of Lemma \ref{lm:stahl}.

It remains to give the

\be{proof}[Proof of Lemma \ref{lm:stahl}]
Fix $m\geq 1$, a self-adjoint matrix $A$ and a projector $P$ acting on $\C^m$, so that $\ker P=\mathrm{span}\{v\}$ for a normalized vector $v$. We are interested in the eigenvalues of $A-tP$ for large $t>0$, call them $\lambda_1(t),\ldots,\lam_m(t)$. We rewrite the matrix as
$$
A-tP=t(-P+t^{-1}A),
$$
and we let $\mu_1(t),\ldots,\mu_m(t)$ denote the eigenvalues of $(-P+t^{-1}A)$ so that $\lam_j(t)=t\mu_j(t)$. This places the problem in the setting of standard analytic perturbation theory, with $t^{-1}$ being the small parameter. Using the results from chapter 7 ($\S$ 2), of Kato's book \cite{Kato}, it is straightforward to compute the eigenvalues $\mu_j(t)$ to lowest order in $t^{-1}$. (Here we use our assumption that $\ker P$ is one-dimensional, as it essentially reduces the problem to non-degenerate perturbation theory.)

Using $\lam_j(t)=t\mu_j(t)$, we obtain, up to relabeling of the eigenvalues,
$$
\lambda_1(t)=\scp{v}{Av}+o(1),\qquad \lam_{j}(t)=-t+o(t),\quad (j=2,\ldots,m).
$$
Therefore, as $t\to\infty$,
$$
\Tr[\exp(A-tP)]=\sum_{j=1}^m e^{\lam_j(t)}
=e^{\lam_1(t)}+o(1)=e^{\scp{v}{Av}}+o(1).
$$
This proves Lemma \ref{lm:stahl} and thus also Proposition \ref{prop:related} 
\e{proof}

%

\paragraph{Acknowledgments}
The author is grateful to Elliott H.~Lieb for raising the question addressed in this paper. It is a pleasure to thank Mario Berta, Eric Carlen, Rupert L.~Frank and Elliott H.~Lieb for helpful discussions.

\end{document}